\documentclass[%
reprint,  
superscriptaddress,
showpacs,
 amsmath,amssymb,
prl,
]{revtex4-1}

\usepackage{graphicx}
\usepackage{dcolumn}
\usepackage{bm}
\usepackage{todonotes}

\usepackage{color}

\newcommand{\pdiff}[2]{\frac{\partial #1}{\partial #2}}

\newcommand{\new}{\nonumber\\}
\newcommand{\abs}[1]{\left|#1\right|}

\newcommand{\ave}[1]{\left\langle #1 \right\rangle}
\newcommand{\im}{{\rm Im}}
\newcommand{\re}{{\rm Re}}

\newcommand{\PP}{\mathcal{P}}

\usepackage{amsmath}	
\begin{document}

\preprint{APS/123-Qed} \title{ Universal non-mean-field
scaling in the density of state of amorphous solids}

\author{Harukuni Ikeda}
 \email{harukuni.ikeda@lpt.ens.fr}
\affiliation{%
\'Ecole Normale Sup\'erieure, UMR 8549 CNRS, 24 Rue Lhomond, 75005 Paris, France
}

\date{\today}

\begin{abstract} 
Amorphous solids have excess soft modes in addition to the phonon modes
described by the Debye theory. Recent numerical results show that if the
phonon modes are carefully removed, the density of state of the excess
soft modes exhibit universal quartic scaling, independent of the
interaction potential, preparation protocol, and spatial dimensions. We
hereby provide a theoretical framework to describe this universal
scaling behavior. For this purpose, we extend the mean-field theory to
include the effects of finite dimensional fluctuation. Based on a
semi-phenomenological argument, we show that mean-field quadratic
scaling is replaced by the quartic scaling in finite
dimensions. Furthermore, we apply our formalism to explain the pressure
and protocol dependence of the excess soft modes.
\end{abstract}


\maketitle

{\em Introduction.--} The vibrational density of state $D(\omega)$ of
amorphous solid differs dramatically from that of crystals. The
low-frequency modes of crystals are phonons that follow the Debye law
$D(\omega)\sim \omega^{d-1}$, where $d$ denotes the spatial
dimensions~\cite{kittel1996}. On the contrary, $D(\omega)/\omega^{d-1}$
of amorphous solids exhibit a sharp peak at the characteristic frequency
$\omega=\omega_{\rm BP}$, which is referred to as the Boson peak (BP).
This behavior suggests the existence of excess soft modes (ESMs) beyond
that predicted by the Debye
law~\cite{anderson2012amorphous,PhysRevLett.53.2316,malinovsky1986nature}.
For $\omega < \omega_{\rm BP}$, the ESMs are spatially
localized~\cite{taraskin1999anharmonicity,laird1991localized,
mazzacurati1996low,chen2010low,liu2011jamming}. These localized modes
play a central role in controlling the various low-temperature
properties of amorphous solids, such as the specific heat, thermal
conduction, and sound
attenuation~\cite{anderson2012amorphous,PhysRevB.4.2029,anderson1972anomalous,
phillips1972tunneling}. Furthermore, recent numerical studies have
established that the ESMs facilitate the structural relaxation of
supercooled liquids at finite
temperatures~\cite{widmer2006predicting,widmer2008irreversible,lerner2018characteristic},
and the local rearrangement of sheared amorphous solids at low
temperature~\cite{xu2010anharmonic,manning2011vibrational,ding2014soft,ji2018theory,kapteijns2018quick}.

The detailed statistical properties of ESMs have been only recently
investigated via numerical simulations. The ESMs can be separated from
the background phonon modes by using a small size
system~\cite{PhysRevLett.117.035501}, observing the participation
ratio~\cite{mizuno2017continuum}, or introducing
impurities~\cite{marco2015,angelani2018probing}. Remarkably, after
successfully removal of the phonons, the ESMs follow the universal
quartic law $D(\omega)=A_4\omega^4$ for $\omega\ll \omega_{\rm BP}$,
independent of the interaction potentials, preparation protocols and
dimensions~\cite{PhysRevLett.117.035501,wang2018low,kapteijns2018universal}.
Considering the relationship with other physical quantities, it is
important to gain an understanding the mechanism that yields the
$D(\omega)=A_4\omega^4$ law and controls the prefactor $A_4$.

The $d$ independence of the quartic law motivates us to apply mean-field
theory to understand this scaling behavior. The replica theory is now
one of the most mature mean-field theories of amorphous
solids~\cite{castellani2005spin,parisi2010mean,biroli2012random,charbonneau2017glass}.
In particular, near the (un) jamming transition point at which the
system loses rigidity~\cite{o2003jamming,liu2011jamming}, the theory
predicts the exact critical exponents of the contact number and shear
modulus~\cite{degiuli2014effects,charbonneau2017glass}. Furthermore, the
theoretical result of $D(\omega)$ agrees very well with the numerical
results for $\omega>\omega_{\rm BP}$ in $d=2$ and
$3$~\cite{degiuli2014effects,franz2017universality}. The replica theory
predicts that amorphous solids near the jamming transition point are in
the Gardner phase~\cite{charbonneau2017glass,biroli2018liu}, which has
been originally investigated in a class of mean-field spin
glasses~\cite{gardner1985spin,gross1985}. In the Gardner phase, the
density of state has the gapless excitation $D(\omega)\sim\omega^2$ for
$\omega<\omega_{\rm BP}$~\cite{franz2017universality}. However, the
numerical results indicate that $D(\omega)\sim\omega^2$ scaling is
observed only near $\omega\sim\omega_{\rm BP}$, and it is replaced by
$D(\omega)\sim A_4\omega^4$ for $\omega\ll\omega_{\rm
BP}$~\cite{mizuno2017continuum}.

The mean-field replica calculation predicts another source of the
singularity that creates the ESMs, in addition to the trivial phonon
modes.  This singularity is related to the quenching rate, or from a
theoretical perspective, the initial temperature $T_{\rm ini}$ of the
equilibrium supercooled liquid before quenching to produce glass. When
$T_{\rm ini}$ is sufficiently low, the supercooled liquid becomes highly
viscous because of the complex structure of the free-energy landscape
containing multiple minima~\cite{goldstein1969viscous}. After quenching,
the system falls to one of the minima. The minima become gradually
unstable with an increase in temperature and eventually disappear above
the so-called mode coupling transition point $T_{\rm
mct}$~\cite{castellani2005spin,biroli2012random}. This instability
affects the vibrational properties of the zero-temperature amorphous
solids and creates ESMs~\cite{parisi2002euclidean}. This view seems to
be consistent with the numerical result that the excess soft modes close
to $\omega\sim \omega_{\rm BP}$ are indeed enhanced for samples quenched
from higher temperatures~\cite{grigera2003phonon}.  However, the
mean-field prediction, $D(\omega)\sim \omega^2$ for $\omega< \omega_{\rm
BP}$, is again inconsistent with the numerical result where
$D(\omega)\sim A_4\omega^4$ scaling is robustly observed irrespective of
$T_{\rm ini}$~\cite{wang2018low}.

In this Letter, we reconcile the aforementioned discrepancies between
the mean-field replica theory and the numerical results in finite $d$
for small $\omega$ by introducing the effect of the finite $d$
fluctuation to the mean-field density of state in a
semi-phenomenological way. We initially construct a theory to describe
the asymptotic behavior of $D(\omega)$ in high $d$ and show that the
quartic law naturally arises as a consequence of finite $d$
fluctuation. Next, motivated by the $d$ independence of the
$D(\omega)\sim A_4\omega^4$ scaling~\cite{kapteijns2018universal}, we
apply our formalism to explain the numerical results in $d=3$. We show
that our theory well reproduces the correct scaling behavior of the
prefactor $A_4$ near jamming and the $T_{\rm ini}$ dependence of $A_4$
for $T_{\rm ini}\sim T_{\rm mct}$.

{\em Effect of the finite dimensional fluctuation.--} Recent numerical
results confirm that the small $\omega$ behavior of the ESMs,
$D(\omega)\sim A_4\omega^4$, does not depend on the spatial dimensions
$d$~\cite{kapteijns2018universal}. Despite this seemingly mean-field
like behavior, mean-field theory fails to reproduce this quartic law. We
first review the discrepancy between the mean-field and numerical
results in high but finite $d$ and then discuss an approach for solving
this problem.

\begin{figure}[t]
\includegraphics[width=8cm]{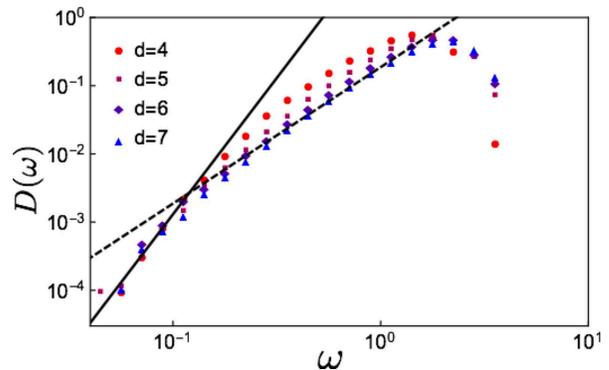} \caption{ Density of state
 $D(\omega)$ of harmonic spheres far from the jamming $\Delta
 \varphi/\varphi_J=1.4$ generated by the fast quench in $d=4$ to $7$. The
 system size is $N=8192$.  Markers denote the numerical result, while
 the dashed and solid lines denote the theoretical predictions,
 $D(\omega)\sim\omega^2$ and $\omega^4$, respectively.  Data are
 reproduced from Ref.~\cite{PhysRevLett.117.045503}.}
 \label{160932_27Nov18}
\end{figure}

In the mean-field replica theory, amorphous solids are modeled by fully
connected models, which are considered to correspond to the $d\to$
limit of the system. The fully connected models have the universal form
of the eigenvalue distribution function for small $\lambda$,
$\rho(\lambda)\sim \sqrt{\lambda-\varepsilon}$~\cite{sm}, where
$\varepsilon$ is proportional to the distance to the instability point.
The mean-field theory predicts that if an amorphous solid is quenched
from high temperature or located near the jamming transition point, the
system becomes marginally stable
$\varepsilon=0$~\cite{franz2015universal,cugliandolo1993analytical},
thus we have
\begin{align}
 \rho_{\rm MF}(\lambda)
 \sim \sqrt{\lambda}\theta(\lambda),
 \label{141431_10Aug18}
\end{align}
where $\theta(x)$ denotes the Heaviside step function. The density of
state $D(\omega)$ is obtained by changing the variable as
$\omega=\sqrt{\lambda}$, which leads to $D(\omega)=2\omega \rho_{\rm
MF}(\lambda=\omega^2)\sim \omega^2$ for small $\omega$. In
Fig.~\ref{160932_27Nov18}, we compare the mean-field prediction with the
numerical result in $d=4$ to $7$. For $\omega>\omega_0\approx 0.12$,
the numerical result converges to the mean-field prediction
$D(\omega)\sim\omega^2$ for an increase of $d$. On the contrary, for
$\omega<\omega_0$, the data systematically deviate from the mean-field
prediction and are well fitted by $D(\omega)\sim\omega^4$, as already
confirmed by previous numerical simulations in $d=3$ and
$4$~\cite{PhysRevLett.117.035501,kapteijns2018universal}. For very small
$\omega$, the numerical results are scattered around $10^{-4}$,
presumably owing to the finite size effect or the lack of statistics
(not shown). We are aware that the size of the current system $N=8192$
is not large enough to observe the effects of
phonons~\cite{kapteijns2018universal}. Nevertheless, we believe that
these effects can be negligible in $d>5$ because the contribution of
the ESMs, $D(\omega)\sim\omega^4$, overwhelms the phonon contribution,
$D(\omega)\sim \omega^{d-1}$.

To clarify the reason for the above discrepancy between the mean-field
and numerical result for finite $d$, we decompose the $i$-th eigenvalue
$\lambda_i$ into the following two parts:
\begin{align}
 \lambda_i = \lambda_i^{\rm MF} + \varepsilon_i,\label{100909_2Dec18}
\end{align}
where $\lambda_i^{\rm MF}$ follows the mean-field result
Eq.~(\ref{141431_10Aug18}) and $\varepsilon_i$ represents the finite $d$
fluctuation. Then, the distribution function of $\lambda_i$ is
\begin{align}
 \rho(\lambda) = \ave{\delta(\lambda-\lambda_i)} =
 \int_{0}^{\infty} d\varepsilon
 \PP_{\lambda}(\varepsilon)\rho_{\rm MF}(\lambda-\varepsilon),\label{202857_26Nov18}
\end{align}
where the lower bound of the integral arises from the stability
condition, $\rho(\lambda)= 0$ for $\lambda<0$, and we introduced the
conditional probability distribution:
\begin{align}
 \PP_\lambda(\varepsilon) =
 \frac{\ave{\delta(\varepsilon-\varepsilon_i)
 \delta(\lambda-\varepsilon-\lambda_i^{\rm MF})}}
 {\ave{\delta(\lambda-\varepsilon-\lambda_i^{\rm MF})}}.
\end{align}
In the $d\to\infty$ limit, the system can be identified with the fully
connected model and thus $\PP_\lambda(\varepsilon)=\delta(\varepsilon)$
to recover the mean-field result. For high but finite $d$,
$\PP_\lambda(\varepsilon)$ is expected to have a narrow distribution
close to $\varepsilon=0$. Thus, we set a small cutoff $\Delta$ and
assume that $\PP_\lambda(\varepsilon)=O(1)$ for $\varepsilon\ll \Delta$
and $\PP_\lambda(\varepsilon)\sim 0$ for $\varepsilon \gg \Delta$. Using
Eqs.~(\ref{141431_10Aug18}) and (\ref{202857_26Nov18}), we obtain the
following for $\lambda\ll\Delta$:
\begin{align}
 \rho(\lambda) 
 &\sim  \PP_0(0) \int_{0}^{\lambda}d\varepsilon
 \sqrt{\lambda-\varepsilon}
\sim  \PP_0(0)\lambda^{3/2},\label{161344_12Aug18}
\end{align}
leading to $D(\omega)\sim \omega^4$.  Thus, the mean-field result is
replaced by the quartic scaling unless the finite dimensional
fluctuation is negligible, \textit{i.e.},
$\PP_\lambda(\varepsilon)=\delta(\varepsilon)$. A similar calculation
leads to $\rho(\lambda)\sim \rho_{\rm MF}(\lambda)$ for
$\lambda\gg\Delta$, meaning that $D(\omega)\sim \omega^2$ for $\omega\gg
\omega_0\equiv \sqrt{\Delta}$. Herewith we recover the numerical results
for high $d$ in Fig.~\ref{160932_27Nov18}.

The $D(\omega)\sim\omega^4$ law is also obtained by a seemingly
different approach: the so-called soft-potential model where the
localized modes are modeled by the collection of anharmonic oscillators
of different
stiffnesses~\cite{ilyin1987parameters,PhysRevB.46.2798,PhysRevLett.89.136801}. The
advantage of our approach over that of the soft-potential model is that
we can consider how the control parameters and preparation protocols
affect the prefactor $A_4$ by relying on the mature replica theory, as
shown in the following sections.

In general, it is impossible to calculate $\PP_\lambda(\varepsilon)$
exactly for finite $d$. To simplify the treatment, we neglect the
$\lambda$ dependence $\PP_\lambda(\varepsilon)\approx \PP(\varepsilon)$,
which is tantamount to neglecting the higher order terms of $\lambda$ and
can be justified for small $\lambda$. Then, Eq.~(\ref{202857_26Nov18})
reduces to
\begin{align}
 \rho(\lambda) = \int_0^\infty d\varepsilon\PP(\varepsilon)
 \rho_{\rm MF}(\lambda-\varepsilon).\label{160419_29Nov18}
\end{align}
From the normalization conditions of $\rho(\lambda)$ and $\rho_{\rm
MF}(\lambda)$, it can be shown that $\int_0^\infty
d\varepsilon\PP(\varepsilon)=1$, suggesting that $\PP(\varepsilon)$ can
be considered as the distribution function of the distance to the
instability point $\varepsilon$. The fluctuation of $\varepsilon$ is a
consequence of the spatial heterogeneity of amorphous solids, which are
not considered in the fully connected mean-field
models~\cite{xia2000fragilities,franz2011field,biroli2014}. The width
$\Delta$ of the distribution $\PP(\varepsilon)$ decreases with an
increase of $d$ as the system approaches the fully connected model. From
the central limit theorem, we expect $\Delta \sim d^{-1/2}$ and the
crossover frequency decreases as $\omega_0\sim d^{-1/4}$.  However, it
is difficult to detect such weak $d$ dependence from the current
numerical result in Fig.~\ref{160932_27Nov18}. Further numerical
investigations are necessary to confirm the $d$ dependence of $\Delta$.

Hereafter, we use a similar argument as that used to derive
Eq.~(\ref{160419_29Nov18}) to analyze the numerical results in $d=3$.
Given that the proposed theory does not taken into account the phonon
mode, the phonon contribution should be removed from the numerical
results as in
Refs.~\cite{PhysRevLett.117.035501,mizuno2017continuum,angelani2018probing},
before comparing with the theoretical prediction.

{\em Pressure dependence near jamming.--} Here we investigate
the pressure $p$ dependence of $D(\omega)$ near the jamming.  For this
purpose, we investigate the negative perceptron model, a mean-field
model of the jamming transition that belongs to the same universality
class of hard/harmonic spheres in the $d\to\infty$
limit~\cite{franz2016simplest,franz2017universality}. The simplicity of
the model allows for analytical calculation of the eigenvalue
distribution function~\cite{franz2015universal}. Near jamming, the model
predicts for $\lambda\ll 1$~\cite{franz2015universal}
\begin{align}
 \rho_{\rm MF}(\lambda) \sim \frac{\sqrt{\lambda}}{\lambda
 + \omega_*^2}\theta(\lambda),\label{164344_2Aug18}
\end{align}
where $\omega_* = c\sqrt{p}$, and $c$ is a constant. Essentially the
same result as Eq.~(\ref{164344_2Aug18}) is obtained by the effective
medium theory, except for the trivial Debye
modes~\cite{degiuli2014effects}. The gapless form of
Eq.~(\ref{164344_2Aug18}) is a consequence of the Gardner
transition~\cite{franz2017universality}, which is the continuous replica
symmetric breaking transition originally discovered in the mean-field
spin glasses~\cite{gardner1985spin,gross1985}. From
Eq.~(\ref{164344_2Aug18}), the scaling behavior of $D(\omega)$ near
jamming ($p\ll 1$) is given as:
\begin{align}
D(\omega)\sim
\begin{cases}
 {\rm constant} & (\omega \gtrsim \omega_*),\\
 (\omega/\omega_*)^2 & (\omega\ll \omega_*).
\end{cases} 
\label{235158_1Aug18} 
\end{align}
However, this is inconsistent with the numerical result in $d=3$. The
numerical result shows that if one carefully removes the phonon modes by
using the participation ratio, one obtains~\cite{mizuno2017continuum}
\begin{align}
 D(\omega)\sim
 \begin{cases}
 {\rm constant} & (\omega \gtrsim \omega_*),\\
  (\omega/\omega_*)^2 & (\omega_0\ll \omega \ll \omega_*),\\
  (\omega/\omega_*)^4 & (\omega\ll\omega_0),
\end{cases}\label{161302_2Aug18} 
\end{align}
where $\omega_0\propto \sqrt{p}$ but the proportionality constant is
much smaller than that of $\omega_*$.

\begin{figure}[t]
\includegraphics[width=8cm]{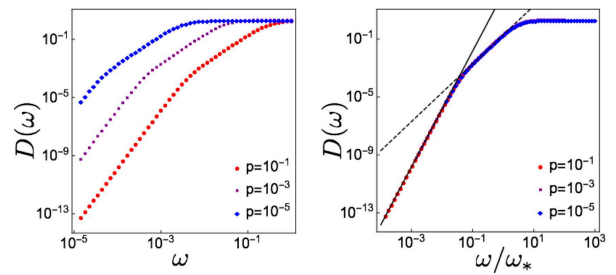} \caption{The density of state
 $D(\omega)$: (left) The results for $p=10^{-1}$, $10^{-3}$ and
 $10^{-5}$. (right) The scaling plot for the same data. The dashed and
 solid lines indicate $D(\omega)\sim\omega^2$ and
 $D(\omega)\sim\omega^4$, respectively.}
				     \label{180755_13Aug18}
\end{figure}

One of the reasons for the discrepancy between the mean-field theory and
the numerical results is the absence of the marginal stability for
finite $d$. The numerical results show that the mean distance to the
instability point has a finite
value~\cite{lerner2014breakdown,degiuli2014effects},
\begin{align}
\ave{\varepsilon} = p\Delta_{\rm G},\label{160206_2Dec18}
\end{align}
where $\Delta_{\rm G}$ is a small positive constant. As in
Eq.~(\ref{160419_29Nov18}), we introduce the fluctuation of
$\varepsilon$ and assume that the mean value of the eigenvalue
distribution function can be written as follows:
\begin{align}
\rho(\lambda)
= \int_0^\infty d\varepsilon \PP(\varepsilon) \rho_{\rm
MF}(\lambda-\varepsilon ).\label{161536_2Dec18}
\end{align}
From Eq.~(\ref{160206_2Dec18}), $\PP(\varepsilon)$ should be
$\PP(\varepsilon) =O(p^{-1}\Delta_{\rm G}^{-1})$ for
$\varepsilon<p\Delta_{\rm G}$ and quickly decreases for
$\varepsilon>p\Delta_{\rm G}$. Repeating a similar argument in
Eq.~(\ref{161344_12Aug18}), we obtain for $\lambda\ll p\Delta_{\rm G}$
\begin{align}
\rho(\lambda) 
 &\sim c^2\Delta_{\rm G}^{-1}\omega_*^{-4}\lambda^{3/2},\label{173412_2Aug18}
\end{align}
which leads to $D(\omega)\sim c^2\Delta_{\rm
G}^{-1}(\omega/\omega_*)^4$.  This scaling smoothly connects to the
mean-field scaling Eq.~(\ref{235158_1Aug18}) at
$\omega\sim\omega_0\equiv \sqrt{p\Delta_{\rm G}}$. Thus, we reproduced
the numerical result, Eq.~(\ref{161302_2Aug18}). Finally, for
concreteness, in Fig.~\ref{180755_13Aug18}, we show $D(\omega)$
calculated by Eq.~(\ref{161536_2Dec18}) assuming $\PP(\varepsilon)=
(p\Delta_{\rm G})^{-1} e^{-(p\Delta_{\rm G})^{-1}\varepsilon}$,
$\Delta_{\rm G}= 10^{-4}$, and $c=1$. If one rescales $\omega$ by
$\omega_*$, the data for different $p$ are collapsed on a single curve
as expected from Eq.~(\ref{161302_2Aug18}).

{\em Initial temperature dependence.--} Here, we discuss the influence
of $T_{\rm ini}$ on the vibrational properties of amorphous solids at
zero temperature. For this purpose, we start from the $p$-spin spherical
model (PSM), which is a prototypical mean-field model for glass
transition to discuss the connection between the glassy slow dynamics
and complex free energy
landscape~\cite{castellani2005spin,biroli2012random}.  The replica
calculation of the PSM shows that there are many metastable states on
the free energy landscape below $T_{\rm mct}$. After quenching, the
system falls to one of the minima.  On the minima, the eigenvalue
distribution function of the PSM follows the Wigner semicircle
law~\cite{kurchan1996phase,biroli2012random}.  For $\lambda\ll 1$ and
$T_{\rm ini}\approx T_{\rm mct}$, this can be written as
\begin{align}
 \rho_{\rm MF}(\lambda) \sim \sqrt{\lambda-\lambda_{\rm min}}\theta(\lambda-\lambda_{\rm min}),
 \label{001533_17Aug18} 
\end{align}
where $\lambda_{\rm min}= c(T_{\rm mct}-T_{\rm ini})$ for $T_{\rm
ini}<T_{\rm mct}$, and $\lambda_{\rm min}=0$ for $T_{\rm ini}\geq T_{\rm
mct}$~\cite{cugliandolo1993analytical}. $c$ is a positive constant.
Repeating a similar argument that used to derive
Eq.~(\ref{160419_29Nov18}), we obtain
\begin{align}
 \rho(\lambda) = \int_{-\lambda_{\rm min}}^{\infty}d\varepsilon
 \PP(\varepsilon)\rho_{\rm MF}(\lambda-\varepsilon),
\end{align}
where the lower bound of the integral is followed by the stability
condition, $\rho(\lambda)=0$ for $\lambda<0$. As mentioned,
$\PP(\varepsilon)$ is expected to have a narrow distribution close to
$\varepsilon=0$.  To express this distribution, we assume
\begin{align}
 \PP(\varepsilon) &\sim
 \Delta_{\rm mct}^{-1} \exp\left[-\abs{\frac{\varepsilon}{\Delta_{\rm mct}}}^\alpha \right],
\end{align}
where $\Delta_{\rm mct}$ and $\alpha$ are constants. Then, the
eigenvalue distribution for small $\lambda$ is calculated as
\begin{align}
 \rho(\lambda) &\sim \PP(-\lambda_{\rm min})\int_{-\lambda_{\rm min}}^{\lambda-\lambda_{\rm min}}d\varepsilon
 \sqrt{\lambda-\lambda_{\rm min}-\varepsilon}\new
 &\sim \PP(-\lambda_{\rm min})\lambda^{3/2},
\end{align}
leading to
\begin{align}
D(\omega) = A_4 \omega^4,\label{001549_17Aug18}
\end{align}
where the prefactor is given as:
\begin{align}
 A_4 &=
 \begin{cases}
  A &  (T_{\rm ini}\geq T_{\rm mct}) \\
  A \exp\left[-\left(\frac{T_{\rm mct}-T_{\rm ini}}{\hat{\Delta}_{\rm mct}}\right)^\alpha \right] &
  (T_{\rm ini}<T_{\rm mct}) 
 \end{cases}.
 \label{153334_12Aug18} 
\end{align}
$A$ is a constant and $\hat{\Delta}_{\rm mct}=\Delta_{\rm mct}/c$. The
preceding equation shows that $A_4$ rapidly decreases for $T_{\rm
mct}-T_{\rm ini}> \hat{\Delta}_{\rm mct}$.
\begin{figure}[t]
\includegraphics[width=8cm]{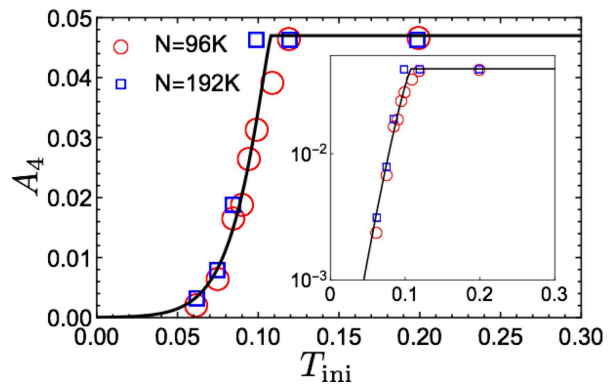} \caption{The $T_{\rm ini}$
dependence of $A_4$ of soft-spheres in $d=3$.  The circles and squares
denote the numerical results of the system sizes $N=96000$ and $192000$,
respectively.
The solid line denotes the theoretical prediction, where $A=0.047$,
 $T_{\rm mct}=0.108$, $\hat{\Delta}_{\rm mct}=0.021$, and $\alpha=1.22$.
 (Inset): The same figure in the semi-log scale. Data are reproduced
from Ref.~\cite{wang2018low}.}  \label{171248_2Aug18}
\end{figure}
In Fig.~\ref{171248_2Aug18}, we compare our prediction with the
numerical results of an amorphous solid in $d=3$.  An excellent
agreement is realized which proves the validity of our theory.

{\em Conclusions and discussions.--} In summary, we discussed the
process by which finite dimensional fluctuation alters the mean-field
scaling $D(\omega)\sim\omega^2$ of amorphous solids.  Our theory
successfully captures quartic scaling, $D(\omega)\sim A_4\omega^4$,
reported in previous numerical simulations. We applied the theory to
describe the pressure $p$ and initial temperature $T_{\rm ini}$
dependence of the prefactor $A_4$. In both cases, the theoretical
results are in good agreement with the previous numerical results.  It
should be noted that the argument in Eq.~(\ref{161344_12Aug18}) does not
depend on the precise form of $\PP_{\lambda}(\varepsilon)$. If
$\PP_\lambda(\varepsilon)$ is finite and continuous at $\varepsilon=0$
and $\lambda=0$, one always obtains the quartic scaling for small
$\omega$. This may explain the robustness of the quartic scaling against
the different interaction potentials, preparation protocol, and
dimensions~\cite{kapteijns2018universal,wang2018low,PhysRevLett.117.035501,
mizuno2017continuum,shimada2018spatial,wang2018low}.

In this Letter, we proposed two singularities to yield the quartic
scaling: singularities related to the Gardner and MCT transitions.  Near
the jamming transition point ($p\ll 1$), one can conclude that the
Gardner transition plays the dominant role in generating quartic scaling
considering the numerical and experimental evidence for the Gardner
transition~\cite{berthier2016growing,jin2017exploring,seguin2016exp} and
the consistency between the mean-field and numerical
results~\cite{charbonneau2014fractal,charbonneau2017glass}. However,
this scenario may not hold apart from jamming where amorphous solids do
not show the strong signature of the Gardner
transition~\cite{scalliet2017absence,seoane2018}. In this region, the
MCT transition would be the main cause of the quartic scaling. For the
intermediate value of $p$, the situation is more complex and further
investigations are needed to determine which singularity plays the
dominant role.

If $\PP_\lambda(\varepsilon)$ is \textit{not} finite at $\varepsilon=0$
and $\lambda=0$, the $D(\omega)\sim \omega^4$ law can be replaced.  For
instance, if $\PP_\lambda(\varepsilon) \sim A\varepsilon^{-\alpha}$ for
small $\varepsilon$ and $\lambda$, a similar calculation as that used in
Eq.~(\ref{161344_12Aug18}) leads to $\rho(\lambda)\sim
A\lambda^{3/2-\alpha}$. This implies that $D(\omega)\sim
A\omega^{4-2\alpha}$. Interestingly, for amorphous solids prepared by
instantaneous quenching without inertia, Lerner and
Bouchbinder~\cite{PhysRevE.96.020104} observed $D(\omega)\sim
\omega^{\beta}$ with $\beta < 4$ suggesting that $\alpha>0$.  Also, some
spring models for amorphous solids on the scale-free
network~\cite{stanifer2018simple} and random graph~\cite{benetti2018}
show that $\beta$ can change depending on the spatial dimensions and
distribution of the coordination number. Further investigations are
required to identify what physical mechanisms control $\alpha$ and
$\beta$.

\begin{acknowledgments}
 {\em Acknowledgments.--} We thank F.~Zamponi, P.~Urbani, A.~Ikeda,
 H.~Mizuno, M.~Wyart, E.~Lerner, W.~Ji, F.~P.~Landes, G.~Biroli,
 L.~Berthier, and G.~Parisi for their participation in useful
 discussions. This project received funding from the European Research
 Coineduncil (ERC) under the European Union's Horizon 2020 research and
 innovation programme (grant agreement n°723955-GlassUniversality).
\end{acknowledgments}

\bibliography{apssamp}

\appendix 
\section{ Supplemental information for ``Universal non-mean-field
scaling in the density of state of amorphous solids''}
Here we briefly explain the scaling behavior of the
mean-field eigenvalue distribution near the minimal eigenvalue. For
more complete and rigorous derivations, see for instance
Refs.~\cite{edwards1976,livan2018introduction}. We consider a Hessian
$H_{ij},\ i,j = 1,\cdots, N$ of a fully connected model. The eigenvalue
distribution function is obtained by using the Edwards and Jones
formula~\cite{edwards1976}:
\begin{align}
 \rho(\lambda) &= \frac{1}{N}\sum_{i=1}^N\delta(\lambda-\lambda_i)\new
 &= \abs{\frac{2}{\pi N}\im \left[\pdiff{}{\lambda}\log Z(\lambda)\right]},\label{160100_8Dec18}
\end{align}
where 
\begin{align}
 Z(\lambda) &= \int d\bm{u} \exp\left[-\frac{1}{2}\bm{u}\cdot\left( H-\lambda I \right)\bm{u}\right]. 
\end{align}
Introducing the new variable $q\equiv \bm{u}\cdot\bm{u}/N$ and using the saddle point method, $Z(\lambda)$ is calculated as
\begin{align}
 \log Z(\lambda) \approx  N\left(\lambda\frac{q}{2}+ G(q)\right),\label{160051_8Dec18}
\end{align}
where
\begin{align}
 G(q) = \frac{1}{N}\log \int d\bm{u} \delta(\bm{u}\cdot\bm{u}-Nq)e^{-\frac{1}{2}\bm{u}\cdot H\cdot\bm{u}}.\label{160930_8Dec18}
\end{align}
For fully connected models, one can directly calculate this quantity by
using the replica method. However, the detailed functional form is not
necessary for our purpose. The saddle point value of $q$ is determined
by
\begin{align}
 \lambda = -2 G'(q).\label{163809_8Dec18}
\end{align}
Substituting Eq.~(\ref{160051_8Dec18}) into Eq.~(\ref{160100_8Dec18}),
we have a simple result:
\begin{align}
 \rho(\lambda) = \abs{\frac{1}{\pi}\im[q(\lambda)]}.\label{162437_8Dec18}
\end{align}
$\rho(\lambda)$ would have a small value near the edge of the
distribution. Eq.~(\ref{162437_8Dec18}) implies that $\im[q(\lambda)]$
is also small near that point. We expand the real and imaginary parts of
Eq.~(\ref{163809_8Dec18}) for $\im[q]$:
\begin{align}
 &\lambda = -2 G'(\re[q(\lambda)]) + \im[q(\lambda)]^2 G'''(\re[q(\lambda)])+\cdots,\new
 &0= \im[q(\lambda)] G''(\re[q(\lambda)])+\cdots.
\end{align}
Then, $\im[q(\lambda)]$ is calculated as
\begin{align} 
\im[q(\lambda)] &\sim\sqrt{\frac{\lambda+2G'(\re[q])}{G'''(\re[q])}}
 \sim  \sqrt{\lambda-\varepsilon},\label{162551_8Dec18}
\end{align}
where $\varepsilon= -2G'(\re[q])$ and $\re[q]$ is determined by
$G''(\re[q])=0$. From Eqs.~(\ref{162437_8Dec18}) and
(\ref{162551_8Dec18}), one can see that $\rho(\lambda)$ shows the square
root singularity near the minimal eigenvalue $\rho(\lambda)\sim
\sqrt{\lambda-\varepsilon}$.

\end{document}